# Document Classification
# Using Distributed Machine Learning

Galip Aydin, Ibrahim Riza Hallac

*Abstract*—In this paper, we investigate the performance and success rates of Naïve Bayes Classification Algorithm for automatic classification of Turkish news into predetermined categories like economy, life, health etc. We use Apache Big Data technologies such as Hadoop, HDFS, Spark and Mahout, and apply these distributed technologies to Machine Learning.

## I. Introduction

Since the number of digital documents grows dramatically, the need for automatic categorization of these documents arises in many different fields. By using machine learning algorithms, documents can be assigned into different categories, titles, languages or even emotional conditions. In this study we describe our work on creating a distributed classification system for collecting the online news and automatically assigning them to related groups using machine-learning algorithms.

## II. Background

### A. Document Classification Steps

Ikonomakis et al. [1] explain the text classification as follows: $d_i$ is a member of a document collection D and $s_j$ is a member of possible categories like $\{s_1, s_2, ..., s_n\}$, then text classification is the operation of matching each $d_i$ with a $s_j$. In this study the classification is done on maximum of five categories. The categories are economy, sports, culture, politics, and world.

Typical steps for applying a machine learning algorithms on text based data are shown in Fig. 1. We followed the same steps in our classification application.

Galip Aydin, Ibrahim Riza Hallac
Firat University
Turkey
gaydin@firat.edu.tr, irhallac@firat.edu.tr

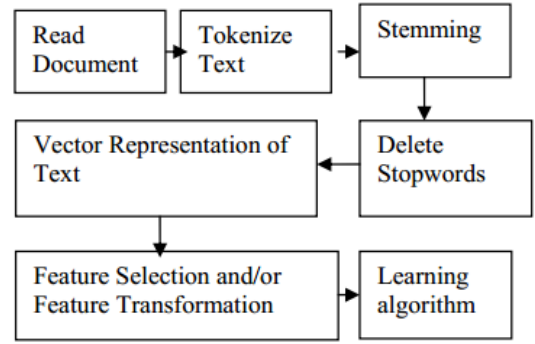

Figure 1. Text classification steps [1]

```
<item>
  <date> date: time </date>
  <category> economy </category>
  <text> the news </text>
</item>
```

The Harvester can be told to run for specific date ranges, categories or news count etc The reason for using the XML is its usefulness in labelling the documents in the training process and gathering the test results.

As shown in the Fig. 1, the first step is reading the documents, which are the XML files created by the News Harvester. The second and third steps (Text tokenization and stemming) require the use of an NLP (Natural Language Processing) toolkit. In our actual implementation the tokenization and stemming processes are done before the document is saved to the disk. Although this was not done in parallel, it still saved a lot of time because we didn't need to re-access the documents later in the training process.

We used a Java based Turkish NLP library called Zemberek [1]. Zemberek is an open source project developed specifically for Turkic languages, especially Turkish. The news contents were divided into sentences and were cleaned from unnecessary words such as stop words other unnecessary parts. In Turkish language words like "ve", "ile", "de", "da" has no meaning alone like "the", "and", "with" words in English. After the tokenization step, we find the root of each word in the stemming step. We only take the word roots into account so that same words are not repeated in the training dictionary.

### B. Data Processing Platform

Unprecedent growth in the size, variety and velocity of data is considered as the Big Data revolution. Google's MapReduce [3] has started a new trend for big data applications. This programming model is not all new but storing and analysing large amounts of data on commodity hardware is gaining a lot of popularity.

Perhaps the most well-known big data framework is Apache's open source distributed and parallel processing framework Hadoop [4]. Hadoop Related projects and solutions keep on coming out really fast.

Another Apache Big Data framework is Spark, which performs 100 times faster than Hadoop [5]. Spark offers an in-memory data processing environment which becomes very useful in iterative computing of big data.

In this study an Apache Spark cluster was set up in Standalone Deploy Mode. Spark version 1.0 was installed on virtual machines. Vector representation of the documents, training and testing was done in parallel on these machines.

## III. Classification

After the pre-processing of the data we need a method to represent the textual data as an algebraic model. Text documents can be considered as arrays of words. They are usually represented as vectors with a weighting factor. This is called as vector space model. In this model there is a dictionary, which consists of all the words of all the documents; the positions of the words are insignificant. The parameters taken into consideration in the weighting of the vector space model are existence of the word, number of each word, total number of the words in the same document and number of the word among all the words in all of the documents.

In this work we used the TF-IDF (term frequency - inverse document frequency) model. The formulation of the TF-IDF method is as follows:

$$M_{ij} = tf_{ij} \tag{1}$$

$$M_{ij} = \log(tf_{ij} + 0.5) * \log\left(\frac{D}{df_{ij}}\right) \tag{2}$$

Here $tf_{ij}$ refers to number of term $i$ in document $j$. $D$ refers to the total number of documents. $df_i$ refers to number of documents contain term $i$.

We calculate the TF-IDF scores in parallel. This process is shown in Fig. 2.

Naive Bayes is a statistical method based on Bayes theory. The possibility of a document $d$ for belonging to the class $s$ can be formulated as:

$$P(s \mid d) = \frac{P(d \mid s) P(s)}{P(d)} \tag{3}$$

One of the pros of the Naive Bayes Classification algorithm is its high success rate even with small training set sizes.

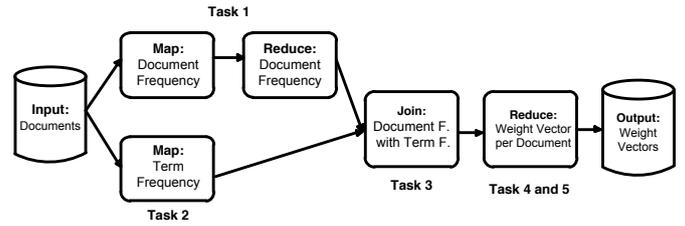

Figure 2. Distributed TF-IDF calculation [6]

## IV. Experimental Results

### A. Classification Results

We implemented different training scenarios to measure the success rate of Naive Bayes Algorithm on news classification.

Our experiment data consists of news from five different categories. We tested the success of classification results for different number of input categories. For each test different number of training data were used. Each training data has to contain at least 20 words. Otherwise the document was ignored.

The experiments contain two steps; first the training is done using a predetermined number of documents for each category, then in the second step 1000 documents are classified using the trained model. The classified documents are not used in the training step.

First we applied the algorithm on economy and sports news. We train the system using various numbers of documents starting from 10 and increasing in each experiment up to 21000. After each training step classification on 1000 documents is done. Then we applied the algorithm on three, four and five categories.

TABLE I. CLASSIFYING TWO CATEGORIES OF NEWS

| Number of training documents per category | Economy Success (%) | Sport Success (%) |
|---|---|---|
| 10 | 79,2 | 88,6 |
| 50 | 96,4 | 94,1 |
| 100 | 96,3 | 97 |
| 250 | 98,4 | 91,6 |
| 500 | 98,8 | 89,4 |
| 1000 | 98,2 | 88,2 |
| 2000 | 97,8 | 87,2 |
| 4000 | 97,6 | 89,1 |
| 8000 | 95,8 | 90,2 |
| 9000 | 96,6 | 90,7 |
| 10000 | 96,7 | 91,2 |
| 18000 | 96,5 | 91,4 |
| 21000 | 95,3 | 92,7 |

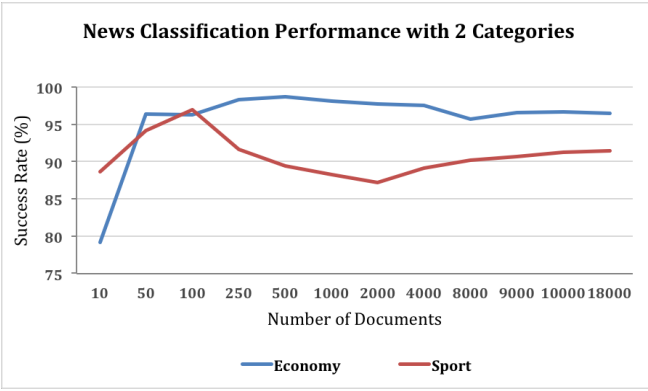

Figure 3.  Classification success rates for two categories

TABLE II. CLASSIFYING FIVE CATEGORIES OF NEWS

| Number of training documents per category | Economy Success (%) | Sport Success (%) | Culture Success (%) | Politics Success (%) | World Success (%) |
|---|---|---|---|---|---|
| 10 | 46,4 | 85,3 | 82,9 | 55 | 53,2 |
| 50 | 88,1 | 89,1 | 92,1 | 58,1 | 85 |
| 100 | 84,2 | 91,6 | 94,1 | 67,8 | 85,6 |
| 500 | 77,2 | 86,9 | 87,7 | 78,5 | 84,7 |
| 1000 | 78,3 | 82,3 | 86,9 | 79,3 | 86,5 |
| 3000 | 78,3 | 80,5 | 83,3 | 81,1 | 84 |
| 5000 | 77,6 | 76,6 | 82 | 79,8 | 80,9 |
| 8000 | 77,4 | 78,1 | 79,2 | 77,4 | 77,8 |
| 10000 | 78,2 | 77,5 | 78,6 | 80,7 | 79,6 |

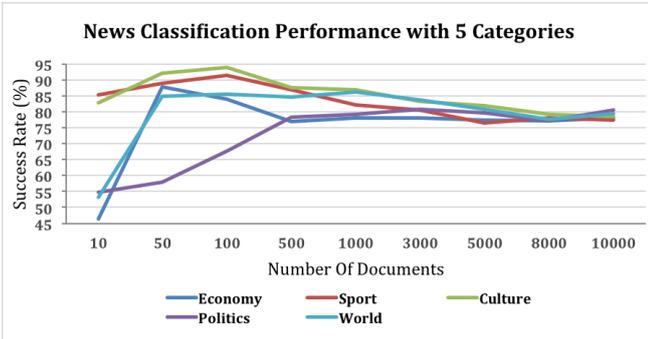

Figure 4.  Classification success rates for five categories

## B. *Performance Tests*

We deployed a private cloud using OpenStack [7] for running the experiments. Spark version 1.0 was installed on virtual machines. Apache Spark provides different cluster manager options. We run the Standalone cluster mode, which is appropriate for quick setup and experimental uses. We used Fabric [8] for managing our cluster. For setting up and running a big data test environment there are some routine tasks for administrating a cluster of nodes and we found Fabric very handy for performing these tasks.

We also measure the effect of the cluster size on distributed classification. To achieve this goal we performed training tests on a single node, two-node and four-node Spark clusters. We change the size of the training set in each step and measure the time. Here training documents consist of same number of labelled documents from four different categories (economy, sports, culture and politics).

Table III shows the running time of training tests on a single node, two-node and four-node clusters for different number of training documents. Each node has 8 GB of RAM. The results show that there is a limit on the number of documents to use for each cluster configuration. The limit is directly related to the amount of RAM available for use by Spark. In single-node configuration the system can not process 32k documents while the two-node cluster can process up to 48k documents. The four-node cluster can successfully process 60k documents in parallel without throwing "out of memory" exception.

We also observed that the number of I/O accesses decrease the performance. In other words using thousands of files as input results in higher processing times. To solve this problem we concatenate hundreds or even a few thousand of files into a larger file, which increases the overall processing performance.

TABLE III. PERFORMANCE OF THE TESTBED

| Number of training documents | Single node Time(sn) | 2 nodes Time(sn) | 4 nodes Time(sn) |
|---|---|---|---|
| 40 | 11 | 11 | 15 |
| 200 | 13 | 13 | 16 |
| 400 | 13 | 13 | 16 |
| 2000 | 17 | 17 | 24 |
| 4000 | 21 | 20 | 29 |
| 8000 | 31 | 27 | 42 |
| 16000 | 48 | 39 | 55 |
| 24000 | 65 | 56 | 69 |
| 32000 | *outOfMem* | 74 | 97 |
| 40000 | *outOfMem* | 88 | 106 |
| 48000 | *outOfMem* | *outOfMem* | 135 |
| 60000 | *outOfMem* | *outOfMem* | 187 |

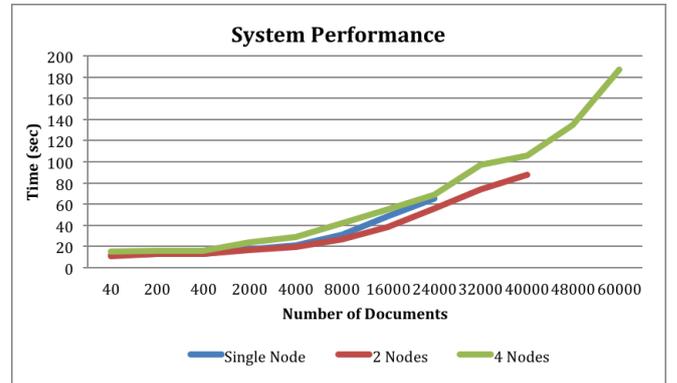

Figure 5.  System Performance Tests

## V. **Conclusions**

In this study we classified Turkish news using Naive Bayes supervised machine learning algorithm. The algorithm was tested with two, three, four and five different categories using up to 50.000 different training documents.

We observed that Naive Bayes performs well even when there is a small number of training documents and it performs better when the training set data grows. When there is large number of documents at hand, it becomes very hard to process all of them in a short time using standard computers. For such situations Big Data technologies may help. We show that using open source cloud computing technologies along with distributed computing frameworks we can process very large number of documents in parallel in very short times.

## *References*

[1] Ikonomakis, M., Kotsiantis, S., and Tampakas, V. *Text classification using machine learning techniques*. WSEAS Transactions on Computers, 2005. 4(8): p. 966-974.

[2] *Zemberek*. [cited 02.01.2015]; Available: https://code.google.com/p/zemberek/.

[3] Dean, Jeffrey, and Sanjay Ghemawat. *MapReduce: simplified data processing on large clusters*. Communications of the ACM 51.1 (2008): 107-113.

[4] *Hadoop*. [cited 02.01.2015]; Available: https://hadoop.apache.org/

[5] ZAHARIA, Matei, et al. *Spark: cluster computing with working sets*. In: Proceedings of the 2nd USENIX conference on Hot topics in cloud computing. 2010. p. 10-10.

[6] *Apache Spark Docs*. [cited 02.01.2015]; Available: http://spark.apache.org/docs

[7] *OpenStack*. [cited 02.01.2015]; Available: http://www.openstack.org/

[8] *Fabric*. [cited 02.01.2015]; Available: http://www.fabfile.org/